# Position Dependent and Independent Evolutionary Models Based on Empirical Amino Acid Substitution Matrices

**Bernardo Barbiellini[1], Alexandra Portnova[2], Anna Chetoukhina[2], Chia-Hsin Lu[2] and Matteo Pellegrini[3]**

[1]Department of Physics
Northeastern University
Boston, MA 02445
Tel: 617-373 2943  Fax:617-373 2943

[2]SIMS, Dept. of Mathematics
Northeastern University
Boston, MA 02445

[3]Department of Chemistry and Biochemistry
University of California
Los Angeles, CA 90095
Tel. 310-825-1402

bba@neu.edu , matteope@mbi.ucla.edu

**Keywords**

Sequence Alignment, Evolution Model, Markov Process, PAM, BLOSUM, amino acid substitution matrix

**Abstract**

Evolutionary models measure the probability of amino acid substitutions occurring over different evolutionary distances. We examine various evolutionary models based on empirically derived amino acid substitution matrices. The models are constructed using the PAM and BLOSUM amino acid substitution matrices. We rescale these matrices by raising them to powers to model substitution patterns that account for different evolutionary distances. We also examine models that account for the dissimilarity of substitution rates along a protein sequence. We compare the models by computing the likelihood of each model across different alignments. We also present a specific example to illustrate the subtle differences in the estimation of evolutionary distance computed using the different models.

## Introduction

Over the past few years a variety of different approaches have been developed to model protein evolution. In general, each model attempts to capture the probability of amino acid changes occurring over a specified time interval during the course of a protein's evolution.  Some of these models start from first principles and account for the rate of substitutions occurring on codons and their effect on amino acid substitutions (Yang 1998).  Others attempt to derive the probability of amino acid substitutions from empirical data obtained by aligning similar proteins (Dayhoff 1978, Henikoff 1992, Muller 2000, Muller 2002).   Although both these approaches have a great deal of merit, here we focus on empirically derived evolutionary models of amino acid substitutions, because the resulting substitution matrices have found widespread use in protein database searches.

Several empirical models that estimate the probability of amino acid substitutions occurring over a specific time have been developed over the past few decades (Dayhoff 1978, Henikoff 1992).  In general, to generate a substitution matrix one must align pairs of sequences that are separated by the same evolutionary distance.  From such alignments it is straightforward to compute the probability of one amino acid being substituted for another over this time.

However, in practice it is difficult to select a priori pairs of protein sequences that are separated by the same evolutionary distance.   Furthermore, to generate the initial alignments between selected pairs one requires a substitution matrix.  Therefore, in this respect, the problem of constructing substitution matrices is circular.

One approach to reduce the impact of the initial substitution matrix is to select only those pairs of sequences that have a small number of substitutions.  Such pairs are easily aligned and the alignments are generally insensitive to the choice of an initial substitution matrix.  Furthermore, for these pairs the evolutionary distances may be estimated approximately.  This approach was used by Dayhoff in the construction of the PAM matrix (Dayhoff 1978).  To construct the matrix she selected pairs of proteins with >85% sequence similarity.   Since these alignments do not depend strongly on the starting amino acid substitution matrix, they minimize the concerns over the circular nature of amino acid substitution matrix construction.

Following the approach proposed by Dayhoff and extended by Gonnet (Gonnet 1992), starting with the initial PAM matrix models of the substitution patterns of amino acids over short evolutionary distances, it is possible to obtain models for longer times by taking powers of this matrix.  Here we assume that evolution is a Markov chain in which previous substitutions do not influence future ones.   Although such an assumption is not accurate over short time scales, it will hold approximately for long evolutionary distances which we are more interested in modeling.  However, the extrapolation of long range substitution patterns from those seen in highly similar sequences will become less accurate as one takes higher powers of the starting matrix.

One strategy to obtain more accurate estimates of substitution patterns over longer evolutionary times is to begin with alignments of more distantly related proteins.  This was the approach taken to construct the BLOSUM amino acid substitution matrix (Henikoff 1992).  This matrix is generated from local, multiple alignments of all commonly occurring motifs in the protein sequence database between sequence pairs whose percent identity was greater than a specified threshold.  Different matrices are constructed using different percent similarity thresholds.

The limitation in using the BLOSUM approach to model amino acid substitutions is twofold.  The sequence alignments used to construct the BLOSUM matrix depend more critically on the initial choice of a substitution matrix than the PAM alignments since the sequence pairs are more divergent. Furthermore the initial pairs are separated by a more heterogeneous set of evolutionary distances than the PAM pairs, so it is less likely that the matrix models substitution over a specific evolutionary distance but rather over a broad distribution of distances.

Regardless of whether one chooses PAM, BLOSUM or other substitution matrices to model evolution, it is important to note that no single matrix will accurately describe substitution rates along an entire sequence since substitutions are strongly position dependent (Yang 1994).   For instance, certain key residues are critical for the activity of an enzyme and their substitution could completely alter the function of a protein while other positions may be substituted with many different amino acids without significantly affecting the protein's function.  During the course of evolution the former position will undergo a very slow rate of substitution, while the latter a much faster one.

If we simply have a pair of aligned sequences it is not possible to reliably estimate which positions in a protein sequences have lower substitution rates than others.  However, in many cases, it is possible to construct a multiple sequence alignment of a protein family.  If the number of sequences aligned is sufficiently large, it is possible to use the amino acid distribution at each position in the multiple sequence alignment to estimate the substitution rates of amino acids.  It is then possible to derive a different evolutionary model for each position in the multiple alignment.  The set of these models should in principle more accurately describe the evolution and evolutionary distance between any two members of the sequence family.

In this work we compare amino acid substitution models constructed using PAM and BLOSUM in an attempt to discover which of these produces more accurate models of evolution over a variety of distances.  For both cases we compare the model described by a single matrix to ones in which we adopt a different matrix for each position along a protein's sequence.  We evaluate the success of the model by comparing the probability of obtaining the amino acid substitution patterns found across alignments for the different models.

## Method

**Evolutionary Models of Amino Acid Substitutions**

The evolution between two protein sequences can be modeled by a Markov process with substitution rates over an evolutionary distance given by a conditional probability matrix: $M_{ji} = p(i \rightarrow j)$, where $p(i \rightarrow j)$ is the probability that amino acid $i$ will be replaced by amino acid $j$. Once we have a transition probability matrix $M$, for each aligned pair of amino acid sequences of length $n$ we can evaluate the probability of one sequence mutating into another as:

(1)    $P(p) = \prod_{n} p(i_n \rightarrow j_n)$,

where $p = p^1$ is the transition probability matrix corresponding to one distance unit.

Evolutionary models for different evolutionary distances can be generated by taking powers of the conditional probability matrix: $p' = p^{\alpha}(i \rightarrow j)$ (note that the powers $\alpha$ need not be integers but can be any real number). The power $\alpha$ that maximizes $P$ is defined to be the optimal model to describe the amino acid substitutions between two aligned sequences. The exponent, however, may not be easily interpreted as some unit of time since the rate at which substitutions occur may be very different during different stages of evolution. This is in part due to the fact that substitution rates are affected by reproductive cycle times which vary greatly from single celled to multi celled organisms.

In real applications of evolutionary models, however, we are limited to an analysis of modern sequences and do not usually know the sequence of the ancestral protein. Therefore it would be incorrect to model the evolution of one modern sequence into another. The correct assumption is that both sequences evolved from an unknown common ancestor. To identify the optimal evolutionary model for this scenario we maximize the log-odds of the joint distribution of two aligned amino acids divided by their expected background distribution which gives us the ratio of the probability that the two sequences derived from a common ancestor versus the null model that they evolved independently of each other (see Appendix).

We obtain a log-odds scoring matrix from the transition probabilities using the relationship:

(2) $\quad S = 10 \log \left( \dfrac{p(j \rightarrow i)}{f_i} \right) = 10 \log \left( \dfrac{M_{ij}}{f_i} \right),$

where $f_i$ is the abundance of amino acid $i$ and we have defined $M_{ij}$ to be the matrix of transition probabilities from amino acid $j$ to $i$. Note that since $M^{\infty} f = f$, $f$ is an eigenvector of $M$

(3) $\quad Mf = f$

and the matrix with elements $\dfrac{M_{ij}}{f_i}$ is symmetric (i.e. $\dfrac{M_{ij}}{f_i} = \dfrac{M_{ji}}{f_j}$) (Gonnet 2002).

Using dynamic programming (Neeldeman 1971, Smith 1981) it is possible to find the alignment between two sequences that generates the maximum sum of the log-odds scores:

(4) $\quad P(S) = 10 \log \prod_n \dfrac{p(i_n \rightarrow j_n)}{f_{j_n}} = \sum_n S_{i_n j_n}$

where $n$ are the positions along an alignment. Maximizing this log-odds score for all aligned positions between two evolutionarily related proteins corresponds to maximizing the probability that the two sequences descended from a common ancestor.

By taking powers of the transition matrix $M$ and converting them to log-odds scores using equation 2, we generate a family of substitution matrices that describe different evolutionary distances. We wish to establish which of these matrices best describes a particular alignment between two proteins. To accomplish this we identify the maximal log-odds score for each alignment using one of these matrices, and then select the matrix that generates the highest maximal log-odds score:

(5) $$P(S(\alpha)) = 10\log\prod_n \frac{p^\alpha(i_n \to j_n)}{f_{j_n}} = \sum_n S_{i_n j_n}(\alpha)$$

The matrix that generates the highest log odds score is considered to be the optimal evolutionary model. The value of $\alpha$ is a measure of the evolutionary distance between the two proteins.

**Evolutionary models derived from PAM**

To compute evolutionary models using the PAM matrix we start with the transition count matrix $M =$ PAM1 shown in Table I (Dayhoff 1978, Gonnet 2002). Since these transitions are observed from sequences with high similarity we see that the diagonal entries $M_{ii}$ are much larger than the off-diagonal entries $M_{ij}$. From this matrix we compute the vector of distributions of amino acids, i.e. the eigenvector of $M$ with eigenvalue 1, $f$ (see equation 3).

Following previous conventions, we raise the $M$ matrix to a power that corresponds to a 1% change in the amino acids between aligned sequences. This percent change in amino acids is computed using the formula (Benner 1994):

(9) $$Diff = 1 - \sum_{i=1}^{20} f_i M_{ii}$$

The matrix that generates a 1% change is defined to be the PAM1 matrix. All successive matrices are generated by taking powers of this matrix.

For each power, $\alpha$, we calculate a matrix $D(\alpha)$, with elements

(5) $$D_{ij}(\alpha) = \frac{M_{ij}^\alpha}{f_i}.$$

We note that $D$ is symmetric. Following previous conventions, we then generate the scoring matrix: $S = 10\log D$.

**Evolutionary models derived from BLOSUM62**

To compute an evolutionary model based on the BLOSUM matrices we start with the BLOSUM62 log-odds scoring matrix that is reported in the literature and shown in Table II (Henikoff 1992). Let us denote it by $B62$. In order to generate a family of evolutionary models using this matrix, we must first convert the log-odds matrix to a transition matrix. To accomplish this we note that the elements of this matrix are defined as

(6) $$B62_{ij} = 2\log_2(\frac{q_{ij}}{f_i f_j}),$$

where $q_{ij}$ is the frequency of amino aligned amino acid pairs $i$ and $j$ observed in all examined sequences in the BLOCKs database (since we are working with BLOSUM62, sequences with 62% and higher similarity were considered) and $f_i f_j$ is the expected frequency of occurrence of a given pair computed from the amino acid abundances $f_i$. Thus an element of $B62_{ij} = 0$ if the observed number of occurrences of aligned pairs of amino acids is equal to the expected frequency of unaligned pairs.

From this log-odds matrix we obtain an odds matrix A with elements

(7) $\quad A_{ij} = 2^{\frac{B62_{ij}}{2}} = \frac{q_{ij}}{f_i f_j}.$

Now, $M_{ij} = f_i A_{ij}$, where $M_{ij}$ is our conditional probability that amino acid $j$ will be replaced by amino acid $i$ through point mutations according to BLOSUM62 and $f_i$ is the abundance of amino acid $i$.

Note, that a priori the $f_i$'s are unknown. To compute them we recall that they must satisfy normalization conditions:

(8) $\quad \sum_i M_{ij} = 1$, or $\sum_i f_i A_{ij} = 1$.

Thus, we obtain a system of 20 linear equations giving as a solution vector $f$. This system of linear equations may be expressed as $A \cdot f = y$, where $y$ is a 20 x 1 vector of 1's. Note that $A$ is symmetric and thus $A = A'$. It follows that $f = A^{-1} y$.

Once we know the vector $f$, we can calculate elements of $M$. We then adjust this matrix $M$ by finding the exponent that applied to $M$ will result in the same percent amino acid change as PAM1, which is 1%. We found that the exponent is 0.078. We next take powers of the adjusted BLOSUM transition matrix to generate a family of evolutionary models. As in the case of the PAM matrices, we finally convert these transition matrices to log odds matrices.

**A Position Specific Evolutionary Model**

In many cases it is possible to go beyond pair-wise protein sequence alignments to generate multiple alignments of a protein family. Unlike in the case of sequence pairs, where dynamic programming can identify the optimal alignment that maximizes the log-odds score (Needleman 1970, Smith 1981), in the case of multiple alignments it is not possible to find the optimal solution which maximizes the sum of all pair-wise scores. However, many approximate solutions to this problem have been developed that find near optimal alignments. One popular program for generating multiple alignments is CLUSTALW (Higgins 1994).

Once the multiple alignment is constructed it is possible to measure the substitution rates for each position in the alignment. Typically, we observe that the substitution rates may change dramatically between different positions. For example, in Table III we see the multiple alignment for Triosephosphate isomerase. Position 2 contains only asparagines (N) while position 19 contains lysine, proline, threonine and glutamine (KPTQ). Therefore, a comprehensive model of the evolution of Triosephosphate isomerase should account for this variability, rather than using a single substitution matrix for all positions.

To generate a position specific evolutionary model, we utilize a different scoring matrix for each position of a multiple alignment. The matrix is selected from the family of matrices that is generated by raising the PAM or BLOSUM matrices to different powers, as explained above. To identify which of these matrices is optimal for a specific position in a multiple alignment, we compare the sum of the score of all possible pair-wise combinations of amino acids at a specific position in the multiple alignment (a column of Table III),

(10)  $$P(S(\alpha,n)) = 10\log \prod_m \frac{p^\alpha(i_n \to j_n)}{f_{j_n}} = \sum_m S_{i_n j_n}(\alpha)$$

where we are taking the product and sum over the $N(N-1)/2$ possible pairs of sequences in the multiple alignment at position $n$, and select the matrix that generates the highest score.

We can think of the set of matrices for each position in the multiple alignment as a vector of α's, with each α representing a different evolutionary model for that position. The values of α for the multiple alignment of Triosephosphate isomerase are reported in Table III. We note that positions with low values of α are highly conserved and thus may participate in the conserved sequence motifs of active sites. In contrast, highly variable positions have large values of α, and may represnt loops or other functionally less important residues.

In the ideal case all the sequences in the multiple alignment would be evolutionarily equidistant from each other so that a single matrix could accurately capture their substitution patterns at each position. However, in practice we find a distribution of evolutionary distances. To select the optimal matrix for a position one could in principle select a subset of the sequences in the multiple alignment that minimize the variance between their evolutionary distances. However, for simplicity, in the current approach we retain all the sequences and therefore our attempt to capture the evolution at each position in the alignment with a single matrix is only an approximation. Nonetheless, we show below that this approximation improves the overall evolutionary model for the family, since neglecting the variability in substitution rates across a sequence is an even more severe approximation.

Once we have computed the optimal substitution matrix for each position in the multiple alignment we can compute the optimal evolutionary model between any two sequences by computing the fraction of the α vector that maximizes the overall alignment:

(11)  $$P(\lambda) = \sum_n S_{i_n j_n}(\lambda \alpha_n)$$

where $\alpha_n$ is the power computed for the $n^{th}$ position of the alignment and λ is the fraction of these exponents that maximizes the log-odds ratio probability. The fraction λ now replaces the coefficient α that was used before to select the optimal evolutionary model. Thus the value of λ may be viewed as an evolutionary distance between the components of a multiple sequence alignment. Of course, the values of λ found for different multiple alignments cannot be directly compared, since they are computed for different protein family specific evolutionary models.

**Results and Discussions**

As a first test case, we consider a multiple alignment consisting of five sequences of Triosephosphate isomerase of which four are almost identical sequences and one is more divergent (see Table III). The sequences are from human, monkey, spinach, mosquito and rice.

We compute the optimal evolutionary model between theses sequences using the PAM and the BLOSUM based approaches. The results are shown in Table IV, where we tabulate the evolutionary distances (values of the exponent) and the sum of the log-odds scores for each protein pair using both matrices. We see that that the PAM and BLOSUM matrices yield roughly equivalent evolutionary distances. However the distances do differ slightly between distant sequences. In general we observe that the BLOSUM method tends to give shorter distances.

These results are in accordance with our expectations that the BLOSUM matrices reduce distances between distant homologous. This is probably due to the fact that, as discussed above, the BLOSUM matrix is constructed by aligning more distant sequences than those used to construct the PAM matrices. However, in general this example demonstrates that the differences between the two matrices are subtle, and that the evolutionary model constructed using BLOSUM is not significantly different from a model based on PAM.

The fact that the two evolutionary models generate slightly different distances leads us to ask the question: which model is generally more accurate? To answer this question we may use the sum of the log-odds scores computed using both approaches. This sum is an estimate of which model generates higher probabilities that pairs of sequences in the multiple alignment are descended from a common ancestor. Since we know a priori that the sequences are evolutionarily related, due to the similarity in their sequences, the model that generates the higher probability is necessarily the more accurate one.

We see in TABLE IV that in all cases but one (rice versus mosquito) the BLOSUM matrix generates higher sums of the log-odds ratios than the PAM approach. This result indicates that the evolutionary models based on BLOSUM more accurately model this multiple sequence alignment than the PAM matrices.

The above calculations were performed using the same substitution matrix for each position in a sequence. If, as described above, we expand the model, and use a different substitution matrix for each position of a multiple sequence alignment, we obtain the results summarized in Table V. In this example we have again computed the distances between aligned fragments of the sequences of the protein triosephosphate isomerase and reported both the evolutionary distances and the sum of the log-odds scores. In this case the evolutionary distances are measured in terms of the $\lambda$ parameter described in equation 11.

We cannot directly compare the evolutionary distances generated by the position dependent models to the position independent ones, since the models are qualitatively different. However, we can compare the ratio of differences between distances using these two approaches. For instance, the mosquito to monkey distances computed using the position independent BLOSUM model is 115, while that for human to monkey is 4. The ratio of these two distances is 28.75, which means that a mosquito is 28.75 times more distant from a monkey than a human is to a monkey. If we compute the same ratio using the position dependent BLOSUM matrix we obtain 40.

When we compute similar ratios of distances between pairs of organisms we observe that the position dependent approach generally generates larger ratios (1.28 vs 1.25 for mosquito to monkey divided by spinach to monkey and 1.86 vs 1.77 for spinach to monkey divided by rice to monkey). In other words, the position dependent approach seems to dilate the distances between species in a nonlinear fashion with respect to the position dependent approach. This may mean that the position dependent approach can resolve these differences with greater accuracy.

To confirm this interpretation we again look at the sum of the log-odds scores. Unlike the interpretation of the distances, the sum of the log-odds may be directly compared between the different approaches, since they are always computing the same quantity (see equation 4). We note in Table V that once again the BLOSUM matrices outperform the PAM matrices. However, we also note that the position dependent model outperforms the position independent ones. Of course this is not entirely surprising since the position dependent model in some sense contains more parameters than the position dependent one. Nonetheless, the differences in log-odds are quite significant (e.g. 56 vs 37 for the mosquito to monkey comparison) and are far more significant than the differences between the PAM and BLOSUM position independent models. Therefore the use of position specific models to describe evolution appears to markedly improve our estimates of whether two sequences are descended from a common ancestor.

We repeated all of the above analyses on a number of other multiple sequence alignments from the BLOCKS (Henikoff 1999) database including ribosomal L6 proteins, S-adenosyl-L-homocysteine hydrolase and bacterial ribonuclease P. We observe that the same trends held as in the multiple alignment of triosephosphate isomerase. In other words, the BLOSUM matrix based models generally performed better than the PAM matrix ones as measured by the sum of the log odds for each pair-wise alignment. Also the BLOSUM matrix models tended to generate smaller distances. Finally, without exception, the position dependent models significantly outperformed the position independent models in all cases.

## Relevance of Approaches and Results

We have compared different models of evolution. The first two models utilize either the PAM or BLOSUM matrices to estimate the amino acid substitutions rates across a protein sequence. In each case we generate an initial transition matrix form the PAM or BLOSUM matrices, and then exponentiate this matrix to model evolution over different distances. In the case of PAM, this is the approach originally proposed by Dayhoff.

Our reason for applying this approach using the BLOSUM matrix is that this matrix is constructed from the substitution rates observed in alignments between more distant proteins than the PAM matrix. We might therefore expect it to generate more reliable evolutionary models between distant homologs.

In the example discussed above, where we analyze triosephosphate isomerase proteins from different organisms, we show that although the distances measured between PAM and BLOSUM matrices are similar, subtle differences do emerge. In particular, we noted that the PAM matrix likely overestimates the distance between distant homologs. Perhaps a more significant finding is that the BLOSUM models are in general more accurate. The accuracy is estimated by comparing the sum of the log-odds ratios generated by the two approaches.

We have also proposed a position specific model of evolution in order to account for the dramatic differences in mutation rates between different positions in a protein sequence. We have applied the position specific calculation to the same family of proteins and shown that it increases the calculated distances between remote homologs with respect to those between close homologs compared to the single matrix approach. We believe that this is evidence that the position specific methodology has increased sensitivity, with respect to the single matrix approach, in the calculation of distances between conserved regions such as motifs and active sites. Once again, we also show that the method generates significantly larger sums of log-odds ratios, and is therefore more accurately modeling the likelihood that the sequences are descended from common ancestors.

The evolutionary models we have considered here only account for substitutions between the twenty amino acids. In the future we would like to expand these models by including the likelihood of insertions and deletions occurring during the course of evolution. Furthermore, we plan to test these evolutionary models against synthetically generated protein families, where the true evolutionary distances are known.

The calculation of accurate evolutionary models is becoming increasingly important in computational biology as the number of proteins being sequenced is dramatically expanding. These models, among other uses, allow one to detect distant homologies and build phylogenetic trees. We believe that the methods that we have presented here, along with other recent innovations that account for variations in amino acid composition across proteins (Yu 2003), bring us closer to the goal of generating simple but accurate models of evolution that could be useful in the interpretation of protein sequence data.

## Appendix

As discussed in Gonnet 2002, to estimate the likelihood that two sequences, *i* and *j*, derive from a common ancestor, *x*, starting with an evolutionary model we compute

$$P(i, j \mid x) = \sum_x f_x P(x \to i) P(x \to j)$$

$$= \sum_x f_x M_{ix} M_{jx}$$

$$= \sum_x M_{ix} f_x M_{jx}$$

$$= \sum_x M_{ix} f_j M_{xj}$$

$$= f_j M_{ij}^2$$

Since the probability of observing an alignment of *i* and *j* if they are not descended from a common ancestor is $f_i f_j$, we obtain the result that the ratio of these probabilities are the log odds values in equation 2.

## Acknowledgements

We thank Prof. Mikhail Malioutov and the participants of the Summer Institute in Mathematical Studies at Northeastern University for clarifying some mathematical aspects of this work, Dr. Nagarajan Sankrithi, Imtiaz Khan and Alper Uzun for developing and testing our coded implementation of the Gonnet's approach. We are also grateful to Rob Henson and Brian Madsen for very useful discussion concerning the Matlab software and its bioinformatics toolbox. We also thank Prof. H. William Detrich for discussions on phylogenetic trees.  This work benefited from the allocation of computer time at the Northeastern University Advanced Scientific Computation Center (ASCC). B.B. was supported by the US Department of Energy under contract DE-AC03-76SF00098.

## References

Benner SA, Cohen MA, Gonnet GH. 1994. Amino acid substitution during functionally constrained divergent evolution of protein sequences. *Protein Eng*., 7(11):1323-32.

Dayhoff MO et al. 1978. *Atlas of Protein Sequence and Structure,* vol. 5 suppl. 3, 345-352.

Gonnet GH, Cohen MA, Benner SA. 1992. Exhaustive matching of the entire protein sequence database. *Science*, 256(5062):1443-5.

Gonnet GH. 2002. Scientific Computation WS 2001/2002 (http://linneus20.ethz.ch:8080/wsrscript.html).

Henikoff S, Henikoff JG. 1992. *Proc. Natl. Acad. Sci. USA* 89:10915-10919.

Henikoff S, Henikoff JG, Pietrokovski S, 1999. Blocks+: A non-redundant database of protein alignment blocks dervied from multiple compilations. *Bioinformatics* 15(6):471-479.

Higgins D, Thompson J, Gibson T, Thompson JD, Higgins DG, Gibson TJ. 1994. CLUSTAL W: improving the sensitivity of progressivemultiple sequence alignment through sequence weighting,position-specific gap penalties and weight matrix choice. *Nucleic Acids Res*., 1994:22:4673-4680.

Muller T, Vingron M. 2000. Modeling amino acid replacement. *J Comput Biol*., 7(6):761-76.

Muller T, Spang R, Vingron M. 2002. Estimating amino acid substitution models: a comparison of Dayhoff's estimator, the resolvent approach and a maximum likelihood method. *Mol Biol Evol*., 19(1):8-13.

Needleman SB, Wunsch CD. 1970. A general method applicable to the search for similarities in the amino acid sequences of two proteins. *J. Mol Bio*. 48:443-453.

Smith TF, Waterman MS. 1981. Identification of common molecular subsequences.  *Journal of Molecular Biology,* 1981:147:195-197.

Yang Z. 1994. Maximum likelihood phylogenetic estimation from DNA sequences with variable rates over sites: approximate methods. *J Mol Evol*., 39(3):306-14.

Yang Z, Nielsen R, Hasegawa M.1998. Models of amino acid substitution and applications to mitochondrial protein evolution. *Mol Biol Evol*., 5(12):1600-11.

Yu YK, Wootton JC, Altschul SF 2003. The compositional adjustment of amino acid substitution matrices.  Proc Natl Acad Sci U S A. 100(26):15688-93.

**Table I: PAM transition matrix (the elements are shown multiplied by 10,000)**

|   | A | R | N | D | C | Q | E | G | H | I | L | K | M | F | P | S | T | W | Y | V |
|---|---|---|---|---|---|---|---|---|---|---|---|---|---|---|---|---|---|---|---|---|
| A | 9890 | 5 | 5 | 6 | 12 | 9 | 11 | 12 | 5 | 2 | 5 | 6 | 9 | 2 | 10 | 29 | 14 | 1 | 2 | 17 |
| R | 4 | 9907 | 5 | 2 | 2 | 16 | 4 | 3 | 8 | 1 | 2 | 30 | 2 | 0 | 3 | 5 | 5 | 4 | 3 | 2 |
| N | 3 | 4 | 9888 | 18 | 2 | 8 | 5 | 6 | 13 | 1 | 1 | 10 | 1 | 1 | 2 | 13 | 8 | 1 | 3 | 1 |
| D | 4 | 2 | 21 | 9905 | 0 | 7 | 28 | 5 | 6 | 0 | 0 | 5 | 0 | 0 | 3 | 7 | 5 | 0 | 1 | 0 |
| C | 3 | 1 | 1 | 0 | 9946 | 0 | 0 | 1 | 1 | 1 | 1 | 0 | 1 | 1 | 0 | 3 | 1 | 1 | 1 | 2 |
| Q | 4 | 11 | 7 | 5 | 1 | 9856 | 18 | 2 | 14 | 1 | 3 | 14 | 6 | 1 | 4 | 5 | 5 | 1 | 1 | 2 |
| E | 8 | 5 | 6 | 30 | 0 | 28 | 9890 | 2 | 7 | 1 | 1 | 15 | 3 | 0 | 4 | 7 | 5 | 1 | 1 | 3 |
| G | 11 | 4 | 9 | 7 | 2 | 4 | 3 | 9952 | 3 | 0 | 1 | 3 | 1 | 0 | 2 | 10 | 2 | 2 | 1 | 1 |
| H | 1 | 4 | 7 | 3 | 1 | 9 | 3 | 1 | 9895 | 1 | 1 | 3 | 2 | 2 | 1 | 2 | 2 | 1 | 9 | 1 |
| I | 2 | 1 | 2 | 0 | 2 | 2 | 1 | 0 | 2 | 9878 | 22 | 2 | 26 | 7 | 1 | 1 | 5 | 2 | 2 | 42 |
| L | 5 | 4 | 2 | 0 | 3 | 8 | 2 | 1 | 3 | 35 | 9919 | 3 | 48 | 22 | 4 | 3 | 4 | 5 | 5 | 19 |
| K | 5 | 33 | 13 | 5 | 0 | 22 | 15 | 2 | 8 | 2 | 2 | 9883 | 5 | 1 | 4 | 6 | 9 | 1 | 2 | 2 |
| M | 3 | 1 | 1 | 0 | 2 | 4 | 1 | 0 | 2 | 10 | 12 | 2 | 9859 | 5 | 0 | 2 | 3 | 1 | 1 | 4 |
| F | 1 | 0 | 1 | 0 | 3 | 1 | 0 | 0 | 4 | 5 | 10 | 0 | 9 | 9923 | 0 | 1 | 1 | 10 | 28 | 3 |
| P | 6 | 2 | 2 | 2 | 0 | 5 | 3 | 1 | 2 | 1 | 2 | 3 | 0 | 1 | 9943 | 6 | 5 | 0 | 1 | 2 |
| S | 23 | 5 | 17 | 8 | 9 | 9 | 7 | 8 | 6 | 1 | 2 | 7 | 4 | 1 | 8 | 9862 | 32 | 2 | 4 | 2 |
| T | 11 | 5 | 11 | 6 | 4 | 8 | 5 | 2 | 7 | 6 | 2 | 9 | 7 | 2 | 7 | 33 | 9879 | 1 | 2 | 12 |
| W | 0 | 1 | 0 | 0 | 1 | 1 | 0 | 0 | 1 | 0 | 1 | 0 | 1 | 3 | 0 | 0 | 0 | 9956 | 4 | 0 |
| Y | 1 | 2 | 2 | 1 | 3 | 1 | 1 | 0 | 13 | 1 | 2 | 1 | 2 | 22 | 1 | 2 | 1 | 10 | 9924 | 2 |
| V | 15 | 2 | 1 | 0 | 8 | 3 | 4 | 1 | 2 | 51 | 14 | 3 | 12 | 5 | 2 | 3 | 14 | 1 | 4 | 9884 |

**Table II: BLOSUM62 log-odds matrix**

|   | A | R | N | D | C | Q | E | G | H | I | L | K | M | F | P | S | T | W | Y | V |
|---|---|---|---|---|---|---|---|---|---|---|---|---|---|---|---|---|---|---|---|---|
| A | 4 | -1 | -2 | -2 | 0 | -1 | -1 | 0 | -2 | -1 | -1 | -1 | -1 | -2 | -1 | 1 | 0 | -3 | -2 | 0 |
| R | -1 | 5 | 0 | -2 | -3 | 1 | 0 | -2 | 0 | -3 | -2 | 2 | -1 | -3 | -2 | -1 | -1 | -3 | -2 | -3 |
| N | -2 | 0 | 6 | 1 | -3 | 0 | 0 | 0 | 1 | -3 | -3 | 0 | -2 | -3 | -2 | 1 | 0 | -4 | -2 | -3 |
| D | -2 | -2 | 1 | 6 | -3 | 0 | 2 | -1 | -1 | -3 | -4 | -1 | -3 | -3 | -1 | 0 | -1 | -4 | -3 | -3 |
| C | 0 | -3 | -3 | -3 | 9 | -3 | -4 | -3 | -3 | -1 | -1 | -3 | -1 | -2 | -3 | -1 | -1 | -2 | -2 | -1 |
| Q | -1 | 1 | 0 | 0 | -3 | 5 | 2 | -2 | 0 | -3 | -2 | 1 | 0 | -3 | -1 | 0 | -1 | -2 | -1 | -2 |
| E | -1 | 0 | 0 | 2 | -4 | 2 | 5 | -2 | 0 | -3 | -3 | 1 | -2 | -3 | -1 | 0 | -1 | -3 | -2 | -2 |
| G | 0 | -2 | 0 | -1 | -3 | -2 | -2 | 6 | -2 | -4 | -4 | -2 | -3 | -3 | -2 | 0 | -2 | -2 | -3 | -3 |
| H | -2 | 0 | 1 | -1 | -3 | 0 | 0 | -2 | 8 | -3 | -3 | -1 | -2 | -1 | -2 | -1 | -2 | -2 | 2 | -3 |
| I | -1 | -3 | -3 | -3 | -1 | -3 | -3 | -4 | -3 | 4 | 2 | -3 | 1 | 0 | -3 | -2 | -1 | -3 | -1 | 3 |
| L | -1 | -2 | -3 | -4 | -1 | -2 | -3 | -4 | -3 | 2 | 4 | -2 | 2 | 0 | -3 | -2 | -1 | -2 | -1 | 1 |
| K | -1 | 2 | 0 | -1 | -3 | 1 | 1 | -2 | -1 | -3 | -2 | 5 | -1 | -3 | -1 | 0 | -1 | -3 | -2 | -2 |
| M | -1 | -1 | -2 | -3 | -1 | 0 | -2 | -3 | -2 | 1 | 2 | -1 | 5 | 0 | -2 | -1 | -1 | -1 | -1 | 1 |
| F | -2 | -3 | -3 | -3 | -2 | -3 | -3 | -3 | -1 | 0 | 0 | -3 | 0 | 6 | -4 | -2 | -2 | 1 | 3 | -1 |
| P | -1 | -2 | -2 | -1 | -3 | -1 | -1 | -2 | -2 | -3 | -3 | -1 | -2 | -4 | 7 | -1 | -1 | -4 | -3 | -2 |
| S | 1 | -1 | 1 | 0 | -1 | 0 | 0 | 0 | -1 | -2 | -2 | 0 | -1 | -2 | -1 | 4 | 1 | -3 | -2 | -2 |
| T | 0 | -1 | 0 | -1 | -1 | -1 | -1 | -2 | -2 | -1 | -1 | -1 | -1 | -2 | -1 | 1 | 5 | -2 | -2 | 0 |
| W | -3 | -3 | -4 | -4 | -2 | -2 | -3 | -2 | -2 | -3 | -2 | -3 | -1 | 1 | -4 | -3 | -2 | 11 | 2 | -3 |
| Y | -2 | -2 | -2 | -3 | -2 | -1 | -2 | -3 | 2 | -1 | -1 | -2 | -1 | 3 | -3 | -2 | -2 | 2 | 7 | -1 |
| V | 0 | -3 | -3 | -3 | -1 | -2 | -2 | -3 | -3 | 3 | 1 | -2 | 1 | -1 | -2 | -2 | 0 | -3 | -1 | 4 |

**Table III: Multiple Sequence Alignment**

| Position | 1 | | | | | | | | | 10 | | | | | | | | | | 20 | | | |
|---|---|---|---|---|---|---|---|---|---|---|---|---|---|---|---|---|---|---|---|---|---|---|---|
| Monkey | M | N | G | R | K | Q | N | L | G | E | L | I | G | T | L | N | A | A | K | V | P | A | D |
| Human | M | N | G | R | K | Q | S | L | G | E | L | I | G | T | L | N | A | A | K | V | P | A | D |
| Mosquito | M | N | G | D | K | A | S | I | A | D | L | C | K | V | L | T | T | G | P | L | N | A | D |
| Spinach | C | N | G | T | K | E | S | I | T | K | L | V | S | D | L | N | S | A | T | L | E | A | D |
| Rice | C | N | G | T | T | D | Q | V | D | K | I | V | K | I | L | N | E | G | Q | I | A | S | T |
| α values | 56 | 1 | 1 | 97 | 27 | 75 | 47 | 69 | 133 | 67 | 35 | 70 | 114 | 114 | 1 | 27 | 75 | 56 | 86 | 72 | 129 | 25 | 34 |

**TABLE IV: PAM and BLOSUM Position Independent Models**

| | \multicolumn{5}{c}{**PAM Evolutionary Distances (the fractions are multiplied by 100)**} | | | | |
|---|---|---|---|---|---|
| | Monkey | Human | Mosquito | Spinach | Rice |
| Monkey | 0 | | | | |
| Human | 4 | 0 | | | |
| Mosquito | 117 | 104 | 0 | | |
| Spinach | 95 | 85 | 101 | 0 | |
| Rice | 165 | 167 | 165 | 112 | 0 |
| **PAM Scores** | | | | | |
| Monkey | | | | | |
| Human | 168.7213 | | | | |
| Mosquito | 35.4258 | 39.0172 | | | |
| Spinach | 45.4413 | 49.8254 | 38.616 | | |
| Rice | 24.2199 | 23.6724 | 26.8289 | 42.8436 | |
| **BLOSUM Evolutionary Distances (the fractions are multiplied by 100)** | | | | | |
| Monkey | 0 | | | | |
| Human | 4 | 0 | | | |
| Mosquito | 115 | 103 | 0 | | |
| Spinach | 92 | 82 | 96 | 0 | |
| Rice | 163 | 162 | 166 | 100 | 0 |
| **BLOSUM Scores** | | | | | |
| Monkey | | | | | |
| Human | 172.5276 | | | | |
| Mosquito | 37.2071 | 40.8236 | | | |
| Spinach | 50.942 | 55.3751 | 40.2724 | | |
| Rice | 25.0639 | 24.9906 | 25.1766 | 48.0355 | |

**TABLE V: PAM and BLOSUM Position Dependent Models**

| | Monkey | Human | Mosquito | Spinach | Rice |
|---|---|---|---|---|---|
| **PAM Evolutionary Distances** | | | | | |
| Monkey | 0 | | | | |
| Human | 6 | 0 | | | |
| Mosquito | 241 | 213 | 0 | | |
| Spinach | 187 | 172 | 192 | 0 | |
| Rice | 346 | 346 | 402 | 226 | 0 |
| **PAM Scores** | | | | | |
| Monkey | | | | | |
| Human | 168.3152 | | | | |
| Mosquito | 54.3733 | 58.094 | | | |
| Spinach | 64.8316 | 69.4538 | 55.3299 | | |
| Rice | 37.5281 | 36.9763 | 34.3307 | 51.4615 | |
| **BLOSUM Evolutionary Distances** | | | | | |
| Monkey | 0 | | | | |
| Human | 6 | 0 | | | |
| Mosquito | 240 | 215 | 0 | | |
| Spinach | 187 | 166 | 198 | 0 | |
| Rice | 348 | 348 | 399 | 202 | 0 |
| **BLOSUM Scores** | | | | | |
| Monkey | | | | | |
| Human | 172.1837 | | | | |
| Mosquito | 56.7469 | 60.332 | | | |
| Spinach | 69.1631 | 73.7285 | 56.5648 | | |
| Rice | 39.2585 | 39.1759 | 35.497 | 55.4642 | |